\documentclass[preprint,nofootinbib,amsmath,amssymb]{revtex4}

\usepackage[dvips]{graphicx}

\newlength{\figwidth}
\begin{document}

%%%%%%%%%%%%%%%%%%%%%%%%%%%%%%%%%%%%%%%%%%%%%%%%%%%%%%%%%%%%%%%%%%%%%%%%%
\title{A New Approach to Testing Dark Energy Models by Observations}
%%%%%%%%%%%%%%%%%%%%%%%%%%%%%%%%%%%%%%%%%%%%%%%%%%%%%%%%%%%%%%%%%%%%%%%%%

\author{Je-An~Gu}
\email{jagu@ntu.edu.tw} %
\affiliation{Physics Division, National Center for Theoretical Sciences, P.O.\ Box 2-131, Hsinchu, Taiwan, R.O.C.} %
\affiliation{Leung Center for Cosmology and Particle Astrophysics
(LeCosPA), National Taiwan University, Taipei 10617, Taiwan,
R.O.C.}

\author{Chien-Wen Chen}
\affiliation{Department of Physics, National Taiwan University, Taipei 10617, Taiwan, R.O.C.} %

\author{Pisin Chen}
\affiliation{Kavli Institute for Particle Astrophysics and
Cosmology, Stanford Linear Accelerator Center, Stanford
University, Stanford, CA 94305, U.S.A.} %
\affiliation{Department of Physics, Graduate Institute of Astrophysics and
LeCosPA, %
%Leung Center for Cosmology and Particle Astrophysics,
National Taiwan University, Taipei 10617, Taiwan, R.O.C.} %

\begin{abstract}
We propose a new approach to the consistency test of dark energy
models with observations. To test a category of dark energy
models, we suggest introducing a \textit{characteristic} $Q(z)$
that in general varies with the redshift $z$ but in those models
plays the role of a (constant) distinct parameter. Then, by
reconstructing $dQ(z)/dz$ from observational data and comparing it
with zero we can assess the consistency between data and the
models under consideration. For a category of models that passes
the test, we can further constrain the distinct parameter of those
models by reconstructing $Q(z)$ from data. For demonstration, in
this paper we concentrate on quintessence. In particular we
examine the exponential potential and the power-law potential via
a widely used % model-independent
parametrization of the dark energy equation of state, $w(z) = w_0
+ w_a z/(1+z)$, for data analysis.
%%%%%%%%%%%%%%%%%%%%%%%%%%%%%%%%%%%%%%%%%%%%%%%%%%%%%%%%%%%%%%%%%%%%
%We conclude: (1) at the $68\%$ confidence level these two
%potentials are both ruled out; (2) at the $95\%$ confidence level
%the exponential potential remains disfavored, while the power-law
%potential is consistent with data; (3) for the power-law potential
%a negative power between $-2$ and $0$ can fit the data.
%%%%%%%%%%%%%%%%%%%%%%%%%%%%%%%%%%%%%%%%%%%%%%%%%%%%%%%%%%%%%%%%%%%%
This method of the consistency test is particularly efficient
because for all models we invoke the constraint of only a single
parameter space that by choice can be easily accessed.
%%%%%%%%%%%%%%%%%%%%%%%%%%%%% 2009.04 added %%%%%%%%%%%%%%%%%%%%%%%%%%%%%
The general principle of our approach is not limited to dark
energy. It may also be applied to the testing of various
cosmological models and even the models in other fields beyond the
scope of cosmology.
%%%%%%%%%%%%%%%%%%%%%%%%%%%%%%%%%%%%%%%%%%%%%%%%%%%%%%%%%%%%%%%%%%%%%%%%%
\end{abstract}

%%%%%%%%%%%%%%%%%%%%%%%%%%%%%%%%%%%%%%%%%%%%%%%%%%%%%%%%%%%%%%%%%%%%%%%%%%%%%%%%%%%%%%%%%%%%%%%%%%%%%%%%%%%%%%%%%
% 95.36.+x Dark energy (see also 98.80.-k Cosmology)
% 98.80.-k Cosmology (see also section 04 General relativity and gravitation; for origin and evolution of galaxies, see 98.62.Ai; for elementary particle and nuclear processes, see 95.30.Cq; for dark matter, see 95.35.+d; for dark energy, see 95.36.+x; for superclusters and large-scale structure of the Universe, see 98.65.Dx)
% 98.80.Es Observational cosmology
%          (including Hubble constant, distance scale, cosmological constant, early Universe, etc)
%%%%%%%%%%%%%%%%%%%%%%%%%%%%%%%%%%%%%%%%%%%%%%%%%%%%%%%%%%%%%%%%%%%%%%%%%%%%%%%%%%%%%%%%%%%%%%%%%%%%%%%%%%%%%%%%%

\maketitle

%%%%%%%%%%%%%%%%%%%%%%%%%%%%%%%%%%%%%%%%%%%%%%%%%%%%%%%%%%%%%%%%%%%%%%%
\section{Introduction\label{sec:introduction}}
%%%%%%%%%%%%%%%%%%%%%%%%%%%%%%%%%%%%%%%%%%%%%%%%%%%%%%%%%%%%%%%%%%%%%%%

The accelerating expansion of the universe in the present epoch
was discovered in 1998 \cite{Perlmutter:1998np,Riess:1998cb} via
Type Ia supernova (SN Ia) distance measurement. This has been
confirmed by more recent observations, including SN Ia
\cite{SNLS:2005,Riess:2006fw,ESSENCE:2007jb}, cosmic microwave
background (CMB) \cite{WMAP:2008} and large-scale structure (LSS)
\cite{Tegmark:2006} observations. Models with a wide variety of
strategies have been proposed to explain this salient phenomenon.
One of the approaches invokes an energy source, generally referred
to as  ``dark energy'', that provides a significant negative
pressure and therefore a repulsive gravity (anti-gravity).
Examples of dark energy candidates include a positive cosmological
constant \cite{Krauss:1995yb,Ostriker:1995su,Liddle:1996pd} and a
dynamical scalar field such as quintessence
\cite{Caldwell:1998ii,Gu:2001tr,Boyle:2001du} and phantom
\cite{Caldwell:1999ew}.

So far many of these models remain consistent with observational
results. In this situation the dark energy information obtained by
comparing the individual theoretical model with the observational
data is indecisive. Responding to this deficiency, recently
cosmologists attempt to extract the generic features of dark
energy, such as (the constraints on) the equation of state (EoS)
or the energy density as a function of the redshift, % directly
from observational results by invoking model-independent
parametrizations in data analysis
\cite{Linder:2002et,Wang:2004ru,Sahni:2006pa,Daly:2007dn}. It is
hoped that several generic questions can be addressed through this
approach. A particularly important question is: Is dark energy a
cosmological constant? If not, it is essential to explore how the
dark energy density evolves with time. A specific manifestation of
this would be the deviation of the dark energy EoS from $-1$.

Along a similar line with the utilization of a parametrization, in
the present paper we propose a new approach to the testing of the
consistency between observational results and dark energy models
or, more generally, cosmological models. For each category of dark
energy models we suggest introducing a distinct
\textit{characteristic} $Q(z)$ that in general varies with time
and the redshift $z$ but is equivalent to an essential constant
parameter within the scope of that category of models. In general
the quantities, $Q(z)$ and $dQ(z)/dz$, by design can be
reconstructed from data with no reliance on the other parameters
of the models. The consistency between data and models can then be
assessed by comparing the observational constraint on $dQ(z)/dz$
with the theoretical prediction, $dQ(z)/dz = 0$. This approach in
principle provides a simple % and direct
``litmus test'' for each category of dark energy models. In
addition, for a category of models that passes the test we can
further constrain the distinct parameter of those models by
reconstructing $Q(z)$ from data.

Close in spirit to our approach, Zunckel and Clarkson
\cite{Zunckel:2008ti} investigated the consistency test of the
flat $\Lambda$CDM model with a parametrization of the luminosity
distance-redshift relation invoked. For the consistency test of
$\Lambda$CDM, a natural choice of the distinct characteristic is
the dark energy density $\rho_\textsc{de}(z)$ that in general is
time-varying but is constant within $\Lambda$CDM. An equivalent
alternative choice invoked by Zunckel and Clarkson in
\cite{Zunckel:2008ti} is $Q_\Lambda (z) = 1 -
\rho_\textsc{de}(z)/\rho_\textrm{c}$ that within flat $\Lambda$CDM
is equal to the matter density fraction $\Omega_\textrm{m}$,
an essential constant parameter. By comparing the
observational constraint on $dQ_\Lambda (z)/dz$ with zero, one can
assess the consistency between data and the $\Lambda$CDM model.

A conventional method of comparing models with observational
results is the \textit{model-based approach}. In this approach one
directly invokes a specific category of models to fit data and
obtains the constraint on the parameter space associated with
those models (e.g., see
\cite{Barnard:2007ta,Abrahamse:2007te,Bozek:2007ti}). One can then
assess the goodness of fit (e.g., see Ref.~\cite{Davis:2007na})
for manifesting how well a model fits the data. In contrast, the
consistency test is for examining whether the condition required
for a category of models is excluded by observational results,
which is different in spirit from the model-based approach. These
two methods are complimentary to each other in the quest for
revealing the nature of dark energy.

%%%%%%%%%%%%%%%%%%%%%%%%%%%% 2009.04 revised %%%%%%%%%%%%%%%%%%%%%%%%%%%%
Our approach to the consistency test and to the constraining of
dark energy models can be efficiently performed because for all
models we invoke the constraint of only a single parameter space
that by choice can be easily accessed. Our approach is
particularly simple and fast when applied to quintessence. It is
because in this approach one does not need to numerically solve
the coupled field equations, the quintessence field equation and
the Einstein equations. This benefit will be demonstrated in Sec.\
\ref{sec:constrainQ} and discussed in Sec.\ \ref{sec:summary}. In
contrast to the benefit, on the other hand we note that the
utilization of parametrization may be accompanied with a bias
against certain models. For all the approaches invoking
parametrization, the model-(in)dependence and the potential
concomitant bias of the chosen parametrization are a separate and
essential issue that requires further exploration. We are
currently investigating this issue for our approach
\cite{Chen:2009}.
%%%%%%%%%%%%%%%%%%%%%%%%%%%%%%%%%%%%%%%%%%%%%%%%%%%%%%%%%%%%%%%%%%%%%%%%%
%Our approach to the consistency test and to the constraining of
%dark energy models can be efficiently performed. This is because
%for all models we invoke the constraint of only a single parameter
%space that by choice can be easily accessed. In particular, when
%applied to quintessence, our approach is significantly simpler and
%faster than the model-based approach where one needs to
%numerically solve the coupled field equations --- the quintessence
%field equation and the Einstein equations --- repeatedly, which
%can be time-consuming. This benefit will be demonstrated in Sec.\
%\ref{sec:constrainQ} and discussed in Sec.\ \ref{sec:summary}. The
%benefit of our approach partly stems from the utilization of a
%single % model-independent
%parametrization in data analysis for extracting dark energy
%information. We note that for all the approaches invoking
%parametrization the model-(in)dependence of the chosen
%parametrization and the potential concomitant bias against certain
%dark energy models are a separate and essential issue that
%requires further exploration. We are currently investigating this
%issue for our approach \cite{Chen:2009}.
%%%%%%%%%%%%%%%%%%%%%%%%%%%%%%%%%%%%%%%%%%%%%%%%%%%%%%%%%%%%%%%%%%%%%%%%%

To demonstrate our approach, in the rest of the present paper we
will concentrate on quintessence. We attempt to extract physical
information about quintessence from the current observational
results via a widely used % model-independent
parametrization of the EoS of dark energy (DE)
\cite{Chevallier:2000qy,Linder:2002et},
\begin{equation}
w_\textsc{de} (z) \, \textrm{ or } \, w_{\phi}(z) = w_{0} + w_a
(1-a) = w_{0} + w_a z/(1+z) \, . \label{eq:w0wa-parametrization}
\end{equation}
Here $z$ is the redshift and $a$ the scale factor of the universe;
the present scale factor $a_0$ has been set to unity by assuming
the (spatial) flatness of the universe for simplicity. The
constraints on $w_{0}$ and $w_a$ from observational data have been
obtained by Riess et al.\ \cite{Riess:2006fw}. We adopt the
constraints of these two parameters with the ``weak prior''
\cite{Riess:2006fw}, for which they invoked the SN Ia data
\cite{Riess:2006fw} as well as the constraints from the LSS
measurement \cite{Tegmark:2003uf,Cole:2005sx}, the baryon acoustic
oscillation (BAO) measurement \cite{Eisenstein:2005su}, the
Cepheid measurement \cite{Freedman:2000cf,Riess:2005zi} and the
3-year WMAP results for CMB \cite{WMAP:2006}. With the
observational constraint on the chosen parameter space we proceed
the testing and the constraining of quintessence in three
directions: (A) We reconstruct the quintessence potential with
data. (B) We assess the consistency between the quintessence
models and data. (C) We constrain the distinct parameter of the
quintessence models by data.

\vspace{1em} \noindent %
\textbf{(A) Reconstruction of potential} \\
To reconstruct the quintessence potential, we will convert the
constraint on the chosen parameter space $(w_0,w_a)$ to that on
the potential $V(\phi)$. The results will be presented in Sec.\
\ref{sec:III-A}.

%%%%%%%%%%%%%%%%%%%%%%%%%%%% 2009.04 revised %%%%%%%%%%%%%%%%%%%%%%%%%%%%
Quintessence can be reconstructed
\cite{Huterer:1998qv,Saini:1999ba,Gerke:2002sx} from data via the
parametrization of its potential \cite{Sahlen:2006dn} or other
relevant physical quantities. In the present demonstration of our
approach we invoke the parametrization of the dark energy EoS
(e.g., see \cite{Guo:2005ata}) in Eq.
(\ref{eq:w0wa-parametrization}). The reconstruction in this way is
particularly simple when the observational constraints on the EoS
parameters can be easily accessed or have already been obtained.
%%%%%%%%%%%%%%%%%%%%%%%%%%%%%%%%%%%%%%%%%%%%%%%%%%%%%%%%%%%%%%%%%%%%%%%%%

\vspace{1em} \noindent %
\textbf{(B) Assessment of consistency} \\
To assess the consistency between quintessence models and data,
for each category of quintessence models we invoke a
characteristic $Q(z)$ that possesses the following features, as
our guidelines of constructing $Q(z)$.
%%%%%%%%%%%%%%%%%%%%%%%%%%%%%%%%%%%%%%%%%%%%%%%%%%%%%%%%%%%%%%%%%%%%%%%%%
\begin{enumerate}
\vspace{-0.2em} \item In general $Q(z)$ is time-varying. %
\vspace{-0.5em} \item Within the scope of the models under consideration $Q(z)$ is constant. %
\vspace{-0.5em} \item For those models $Q(z)$ plays the role of a distinct parameter. \\ %
(This feature is oriented to fulfilling Direction C.)
\vspace{-0.5em} \item By our construction $Q(z)$ is a functional
of the parametrized physical quantity $P(z)$, which in the present
demonstration is the dark energy EoS, $w_\textsc{de}(z)$. This is
so designed that $Q(z)$ and its derivative $dQ(z)/dz$ can be
reconstructed from data via the constraint on the parameters
involved in the $P(z)$ parametrization, for example, via the
constraint on $(w_0,w_a)$ as considered in the present
demonstration. %
\vspace{-0.5em} \item Accordingly, the (in)compatibility between the
observational constraint on $dQ(z)/dz$ and the theoretical prediction,
$dQ(z)/dz=0$, tells the (in)consistency between data and models.
\vspace{-0.2em}
\end{enumerate}
%%%%%%%%%%%%%%%%%%%%%%%%%%%%%%%%%%%%%%%%%%%%%%%%%%%%%%%%%%%%%%%%%%%%%%%%%
We will convert the constraint on the chosen parameter space
$(w_0,w_a)$ to that on $dQ(z)/dz$, and compare it with zero,
thereby assessing the consistency between the data and the models
under consideration. The results will be presented in Sec.\
\ref{sec:III-B}.

\vspace{1em} \noindent %
\textbf{(C) Constraining the distinct parameter} \\
For a category of quintessence models with a characteristic $Q(z)$
introduced in the above-mentioned manner, we will convert the
constraint on the chosen parameter space $(w_0,w_a)$ to that on
$Q(z)$, thereby giving the constraint on the distinct parameter of
the potential. The results will be presented in Sec.\
\ref{sec:III-C}.

As a demonstration of the effectiveness of our approach, for
Direction B we examine the exponential potential (for a recent
study see \cite{Bozek:2007ti} and references therein) and the
power-law potential (that with a negative power is invoked in the
tracker quintessence \cite{TrackerQ:1999}).
%%%%%%%
We conclude: (Direction B) at the $68\%$ confidence level both are
ruled out; at the $95\%$ confidence level the exponential
potential remains disfavored, while the power-law potential is
consistent with data. Since the exponential potential is
inconsistent with data, for Direction C we deal with only the
power-law potential for which the power-law index is the distinct
parameter in our treatment. We conclude: (Direction C) for the
power-law potential, at the $95\%$ confidence level a negative
power-law index between $-2$ and $0$ can fit the data, whereas a
positive power is ruled out.

%%%%%%%%%%%%%%%%%%%%%%%%%%%%%%%%%%%%%%%%%%%%%%%%%%%%%%%%%%%%%%%%%%%%%%%
\section{The Basics\label{sec:basics}}
%%%%%%%%%%%%%%%%%%%%%%%%%%%%%%%%%%%%%%%%%%%%%%%%%%%%%%%%%%%%%%%%%%%%%%%

Consider a Friedmann-Lemaitre-Robertson-Walker (FLRW) universe
described by the Robertson-Walker metric,
\begin{equation}
ds^2 = dt^2 - a^2(t) \left( \frac{dr^2}{1-kr^2} + r^2 d \Omega ^2
\right) \, , %
\label{eq:RW metric}
\end{equation}
and assume that it is dominated by pressureless matter and
quintessence in the present epoch. Quintessence is a dynamical
scalar field described by the Lagrangian density,
\begin{equation} \label{eq:scalar Lagrangian}
\mathcal{L} = \sqrt{|g|} \left[ \frac{1}{2} g^{\mu \nu } (\partial
_{\mu} \phi ) (\partial _{\nu} \phi ) - V(\phi ) \right] .
\end{equation}
For simplicity, we assume that the
universe is spatially flat ($k=0$) and that the spatial dependence
of quintessence is weak so that the spatial curvature and the spatial derivative
terms are ignored. The governing equations, i.e.\ the
Einstein equations and the quintessence field equation, for the
cosmic evolution are then as follows.
\begin{eqnarray}
\left( \frac{\dot{a}}{a} \right) ^2 &=& \frac{8 \pi G}{3} \rho =
\frac{8 \pi G}{3} \left( \rho_\textrm{m} + \rho_{\phi} \right) , %
\label{eq:Friedmann eqn} \\
\frac{\ddot{a}}{a} &=& - \frac{4 \pi G}{3} \left( \rho + 3p
\right) = - \frac{4 \pi G}{3} \left( \rho_\textrm{m} + \rho_{\phi} + 3p_{\phi} \right) , %
\label{eq:accel eqn}
\end{eqnarray}
\begin{equation}
\ddot{\phi} + 3H \dot{\phi} + V'(\phi ) = 0 \, , %
\label{eq:phi field eqn}
\end{equation}
where the Hubble expansion rate is defined as $H \equiv
\dot{a}/a$, and the energy density and pressure of quintessence
are given by
\begin{eqnarray}
\rho_{\phi} &=& \frac{1}{2} \dot{\phi}^2 + V(\phi) = K + V \, , \label{eq:phi rho} \\
p_{\phi}    &=& \frac{1}{2} \dot{\phi}^2 - V(\phi) = K - V \, . \label{eq:phi p} %
\end{eqnarray}
The EoS of quintessence is therefore
\begin{equation}
w_{\phi} = p_{\phi} / \rho_{\phi} = (K-V)/(K+V) \, . %
\label{eq:phi w}
\end{equation}

Next we derive the expressions for the relevant quantities in
terms of $w_{\phi}(z)$ and the essential cosmological parameters
such as $\Omega_{\phi}$ and $\Omega_{m}$. From the Einstein
equations, we have
\begin{equation}
\rho_{\phi}(z) = \rho_c \Omega_{\phi} \exp \left( 3 \int_{0}^{z}
\left[ 1 + w_{\phi}(z') \right] \frac{dz'}{1+z'} \right) ,
\label{eq:rho phi (z)}
\end{equation}
where the critical density $\rho_c \equiv 3H_0^2 / 8 \pi G_N$.
Therefore,
\begin{eqnarray}
H^2 (z) &=& \frac{8 \pi G_N}{3} \left( \rho_{m} +
\rho_{\phi} \right) \nonumber \\
&=& H_0^2 \left[ \Omega_{m} (1+z)^3 + \Omega_{\phi} \exp
\left( 3 \int_{0}^{z} \left[ 1 + w_{\phi}(z') \right]
\frac{dz'}{1+z'} \right) \right] . %
\label{eq:H2(z)}
\end{eqnarray}
In addition, from Eqs.\ (\ref{eq:phi rho}) -- (\ref{eq:phi w}), we
have
\begin{eqnarray}
K(z) &=& \left[ 1 + w_{\phi}(z) \right] \rho_{\phi}(z) / 2 \, , \label{eq:K(z)} \\ %
V(z) &=& \left[ 1 - w_{\phi}(z) \right] \rho_{\phi}(z) / 2 \, , \label{eq:V(z)} %
\end{eqnarray}
from the former of which we obtain
\begin{equation}
\phi(z) - \phi_0 = \pm \int_0^z \frac{\sqrt{\left[ 1+w_{\phi}(z')
\right]\rho_{\phi}(z')}}{H(z')} \frac{dz'}{1+z'} \, , %
\label{eq:phi(z)}
\end{equation}
where $\phi_0$ is the present value of the quintessence field.

For a given parametrization of $w_{\phi}(z)$, a parametric
relationship between the potential $V$ and the quintessence field
$\phi$ can be deduced from Eqs.\ (\ref{eq:V(z)}) and
(\ref{eq:phi(z)}) with the help of Eqs.\ (\ref{eq:rho phi (z)})
and (\ref{eq:H2(z)}). Based on this relation, from the
observational constraint of $w_{\phi}(z)$ one can then in
principle reconstruct the quintessence potential $V(\phi)$ as well
as other quantities which can be expressed in terms of $V(z)$,
$\phi(z)$ and their derivatives.

%%%%%%%%%%%%%%%%%%%%%%%%%%%%%%%%%%%%%%%%%%%%%%%%%%%%%%%%%%%%%%%%%%%%%%%
\section{Testing and Constraining Quintessence \label{sec:constrainQ}}
%%%%%%%%%%%%%%%%%%%%%%%%%%%%%%%%%%%%%%%%%%%%%%%%%%%%%%%%%%%%%%%%%%%%%%%

In this section we test and constrain quintessence with the
current observational results via the $w_{\phi}(z)$
parametrization in Eq.\ (\ref{eq:w0wa-parametrization}) in three
directions: (A) reconstructing the quintessence potential
$V(\phi)$, (B) assessing the consistency between a category of
quintessence models and data via the derivative of a distinct
characteristic, $dQ(z)/dz$, and (C) constraining the distinct
parameter of a category of quintessence models played by the
characteristic $Q(z)$. The observational constraints on the two
parameters in $w_{\phi}(z)$ have been obtained by Riess et al.\ in
Ref.~\onlinecite{Riess:2006fw}, as shown in Fig.\ 10 therein. We
invoke the $68\%$ and the $95\%$ confidence contours in the left
panel, i.e.\ with a weak prior, of that figure and focus on
the region satisfying $w(z) > -1 \, \forall \, z$ for quintessence.%
%%%%%%
\footnote{By definition, no quintessence can be reconstructed from
the region where $w<-1$ for some $z$.}
%%%%%%
These contours were deduced from the SN Ia data
\cite{Riess:2006fw} and the constraints from other measurements
\cite{Tegmark:2003uf,Cole:2005sx,Eisenstein:2005su,Freedman:2000cf,Riess:2005zi,WMAP:2006}.
The redshift range of these supernovae is $0<z<1.8$. In the
reconstruction we set $\Omega_{m}=0.28$ and $\Omega_{\phi}=0.72$.

%%%%%%%%%%%%%%%%%%%%%%%%%%%%%%%%%%%%%%%%%%%%%%%%%%%%%%%%%%%%%%%%%%%%%%%
\subsection{Reconstruction of potential}  \label{sec:III-A}
%%%%%%%%%%%%%%%%%%%%%%%%%%%%%%%%%%%%%%%%%%%%%%%%%%%%%%%%%%%%%%%%%%%%%%%

\begin{figure}
\includegraphics{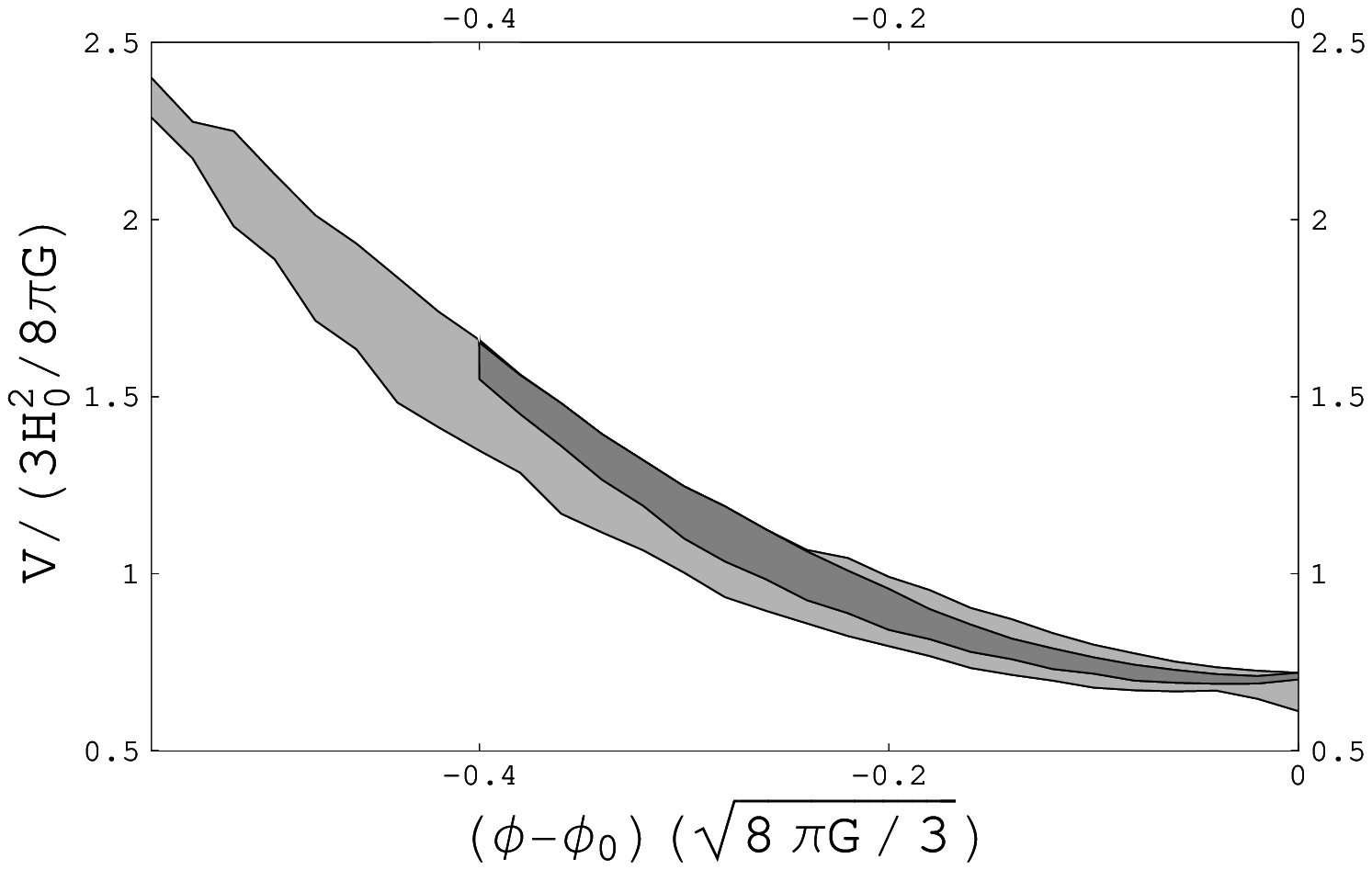}% Here is how to import EPS art
\caption{\label{fig:V-phi} The potential reconstructed from data
via the $w_{\phi}(z)$ parametrization in Eq.\
(\ref{eq:w0wa-parametrization}).}
\end{figure}

The reconstructed potential is sketched in Fig.\ \ref{fig:V-phi},
where the dark gray and the light gray areas correspond to the
$68\%$ and the $95\%$ confidence regions, respectively. From this
figure the shape of the quintessence potential is already
apparent.

Note that the $68\%$ and the $95\%$ confidence contours in the
$(w_0,w_a)$ parameter space, from which the potential is
reconstructed, enclose both the quintessence and the
non-quintessence cases. It is therefore the probability space of
$(w_0,w_a)$, but not that of the quintessence models, that the
confidence of the constraints in Fig.\ \ref{fig:V-phi} is measured
against.

%%%%%%%%%%%%%%%%%%%%%%%%%%%%%%%%%%%%%%%%%%%%%%%%%%%%%%%%%%%%%%%%%%%%%%%
\subsection{Assessment of consistency} \label{sec:III-B}
%%%%%%%%%%%%%%%%%%%%%%%%%%%%%%%%%%%%%%%%%%%%%%%%%%%%%%%%%%%%%%%%%%%%%%%

To assess the consistency between quintessence models and data, we
deal with one category of potentials (in principle embracing
infinitely many specific potentials) at once. We facilitate this
consistency assessment by introducing a {\it characteristic}
$Q(z)$ with several features listed in Introduction. By
construction, $Q(z)$, although is in general time-varying, plays
the role of a constant distinct parameter within the scope of the
models under consideration.

The characteristic $Q(z)$ by design can be expressed in terms of
$V$, $dV/d\phi$, $d^2V/d\phi^2$, $\ldots$.%
%%%%%%
\footnote{For a category of potentials $V(\phi;q_i)$ involving N
parameters, $\{q_i \, , \; i=1,2,\ldots,N\}$, one can treat
$V,V',V'',\ldots,V^{[N]}$ as $N+1$ equalities for $N+1$ variables:
$\phi$ and $q_i$. Then, in general, one can obtain $q_i$ as a
function of $V,V',V'',\ldots,V^{[N]}$ by solving these
equalities.} %
%%%%%%
As shown in the previous section, $V(z)$ and $\phi(z)$ are the
functionals of $w_{\phi}(z)$ and the cosmological parameters $\{
H_0 , \Omega_m , \Omega_{\phi} \}$. Accordingly, the
characteristic $Q(z)$ so constructed and its derivative $dQ(z)/dz$
are also the functionals of $w_{\phi}(z)$ and those cosmological
parameters, and can therefore be reconstructed from data via the
constraint on $w_0$ and $w_a$ (in the same way as that for the
potential reconstruction). Then, by comparing the reconstructed
$dQ/dz$ with zero within the redshift range $0<z<1.8$, one can
determine the consistency between the data and the quintessence
potentials under consideration.

As a demonstration of the effectiveness of this approach, in the
following we will examine the exponential potential and the
power-law potential,
\begin{eqnarray}
V_\textrm{exp}(\phi) &=& V_A \exp \left[ - \phi / M_{0} \right] , \label{eq:exponential potential} \\ %
V_\textrm{power}(\phi) &=& m^{4-n_{0}} \phi^{n_{0}} . \label{eq:power-law potential} %
\end{eqnarray}

To examine the exponential potential in Eq.\ (\ref{eq:exponential
potential}), we introduce the following characteristic,
\begin{equation} \label{eq:Qexp}
Q_\textrm{exp}(z) = M^{-1}(z) \equiv - \frac{1}{V(z)}
\frac{dV}{d\phi}(z) \, ,
\end{equation}
which for the exponential potential is equal to the essential
parameter $1/M_0$. This characteristic, by construction, does not
explicitly involve $\phi(z)$. Naively, to check whether a
(reconstructed) potential follows the exponential behavior in Eq.\
(\ref{eq:exponential potential}) one can plot $\ln V$ versus $\phi
- \phi_0$ and see whether there exists a straight line passing
through the shaded region, for which the characteristic
$Q_\textrm{exp}$ in Eq.\ (\ref{eq:Qexp}) is the slope. Note that
in this exponential case the value of $\phi_0$ does not affect the
existence of the straight line within the shaded region. Instead
of exhausting all straight lines with the try-and-error method, we
take a more efficient approach by comparing the derivative of the
characteristic, $dQ_\textrm{exp}(z)/dz$, with the zero-line (a
single line), as mentioned earlier.

\begin{figure}
\includegraphics{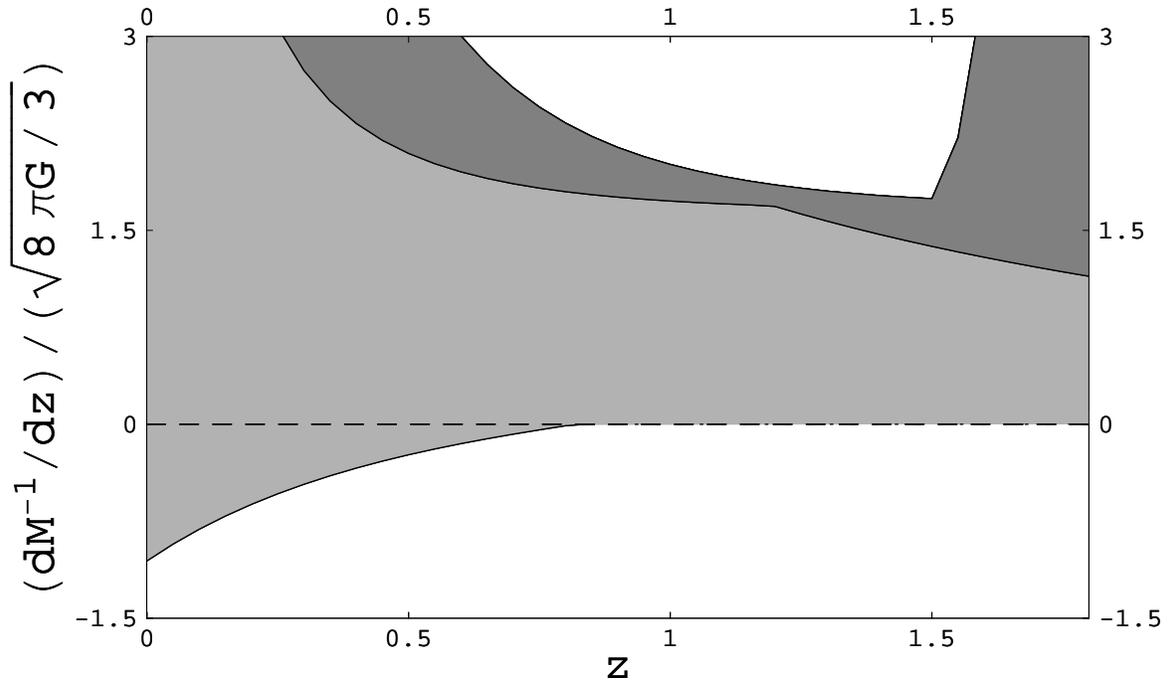}% Here is how to import EPS art
\caption{\label{fig:dM-z} The derivative of the characteristic of
the exponential potential reconstructed from data.}
\end{figure}

The derivative of the characteristic w.r.t.\ the redshift $z$,
$dM^{-1}(z)/dz$, reconstructed from data for $0 < z < 1.8$ is
shown in Fig.\ \ref{fig:dM-z} (dark gray for $68\%$ confidence and
light gray for $95\%$). From this figure, one can see that the
horizontal zero-line lies outside the $68\%$ confidence constraint
for all $z$ and is on the margin of the $95\%$ confidence
constraint for $z>0.8$. Accordingly, at the $68\%$ confidence
level [with regard to the $(w_0,w_a)$ probability space] the
quintessence model with the exponential potential is ruled out,
and at the $95\%$ confidence level the likelihood of this model to
describe the cosmic evolution is marginal for $z>0.8$.

To examine the power-law potential in Eq.\ (\ref{eq:power-law
potential}), we introduce the following characteristic,
\begin{equation} \label{eq:Qpower}
Q_\textrm{power}(z) = n(z) \equiv \left[ 1 - V(z) \left(
\frac{dV}{d\phi}(z) \right)^{-2} \frac{d^2 V}{d\phi^2} (z)
\right]^{-1} ,
\end{equation}
which for the power-law potential is equal to the power-law index
$n_0$, an essential constant that characterizes the potential.
This characteristic, by construction, does not explicitly involve
$\phi(z)$. Naively, to check whether a (reconstructed) potential
follows the power-law behavior in Eq.\ (\ref{eq:power-law
potential}) one can plot $\ln V$ versus $\ln \phi$ and see within
the constrained region whether there exists a straight line, of
which the characteristic $Q_\textrm{power}$ in Eq.\
(\ref{eq:Qpower}) is the slope. However, the variable we
reconstruct is $\phi (z) - \phi_0$, not $\phi(z)$ itself.
Therefore what we can directly plot is $\ln V$ versus $\ln (\phi -
\phi_0)$, but not $\ln V$ versus $\ln \phi$. We note that in this
power-law case the value of $\phi_0$ plays an essential role in
the existence of the straight line, and accordingly in this naive
method one cannot assess the consistency without the information
about $\phi_0$. This is the reason why we avoid invoking $\phi
(z)$ in constructing the characteristics, such as those in Eqs.\
(\ref{eq:Qexp}) and (\ref{eq:Qpower}). Following the same
procedure mentioned in the previous paragraphs, we proceed again
with the derivative of the characteristic.

\begin{figure}
\includegraphics{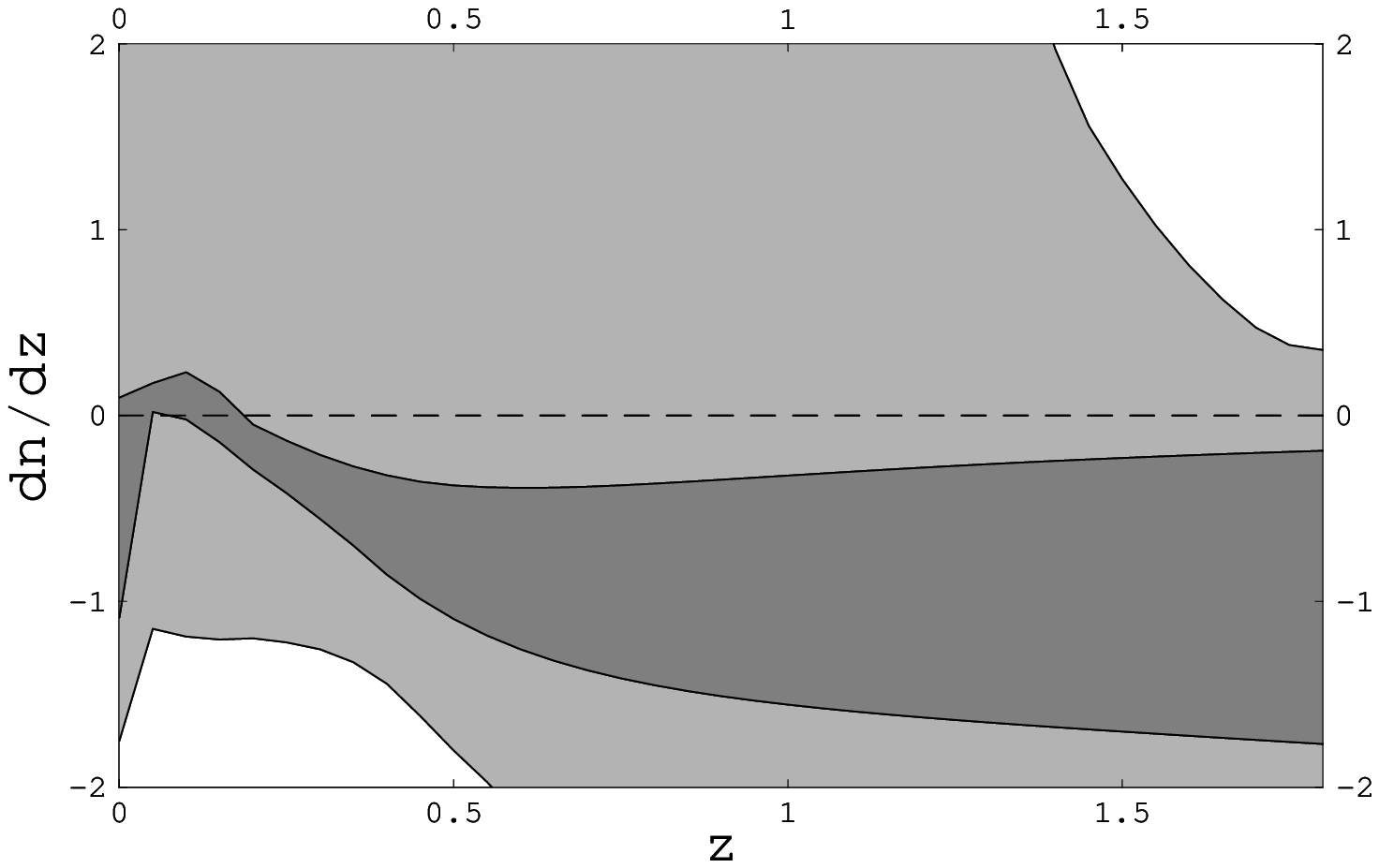}% Here is how to import EPS art
\caption{\label{fig:dn-z} The derivative of the characteristic of
the power-law potential reconstructed from data.}
\end{figure}

The quantity, $dn(z)/dz$, reconstructed from data for $0 < z <
1.8$ is shown in Fig.\ \ref{fig:dn-z} (dark gray for $68\%$
confidence and light gray for $95\%$). As shown in this figure,
the horizontal zero-line is outside the $68\%$ confidence
constraint for most of the redshift $z$ but within the $95\%$
confidence constraint for all $z$ under consideration.
Accordingly, while the quintessence model with the power-law
potential is ruled out at the $68\%$ confidence level, it can fit
the current data at the $95\%$ confidence level [with regard to
the $(w_0,w_a)$ probability space].

%%%%%%%%%%%%%%%%%%%%%%%%%%%%%%%%%%%%%%%%%%%%%%%%%%%%%%%%%%%%%%%%%%%%%%%
\subsection{Constraining the distinct parameter} \label{sec:III-C}
%%%%%%%%%%%%%%%%%%%%%%%%%%%%%%%%%%%%%%%%%%%%%%%%%%%%%%%%%%%%%%%%%%%%%%%

\begin{figure}
\includegraphics{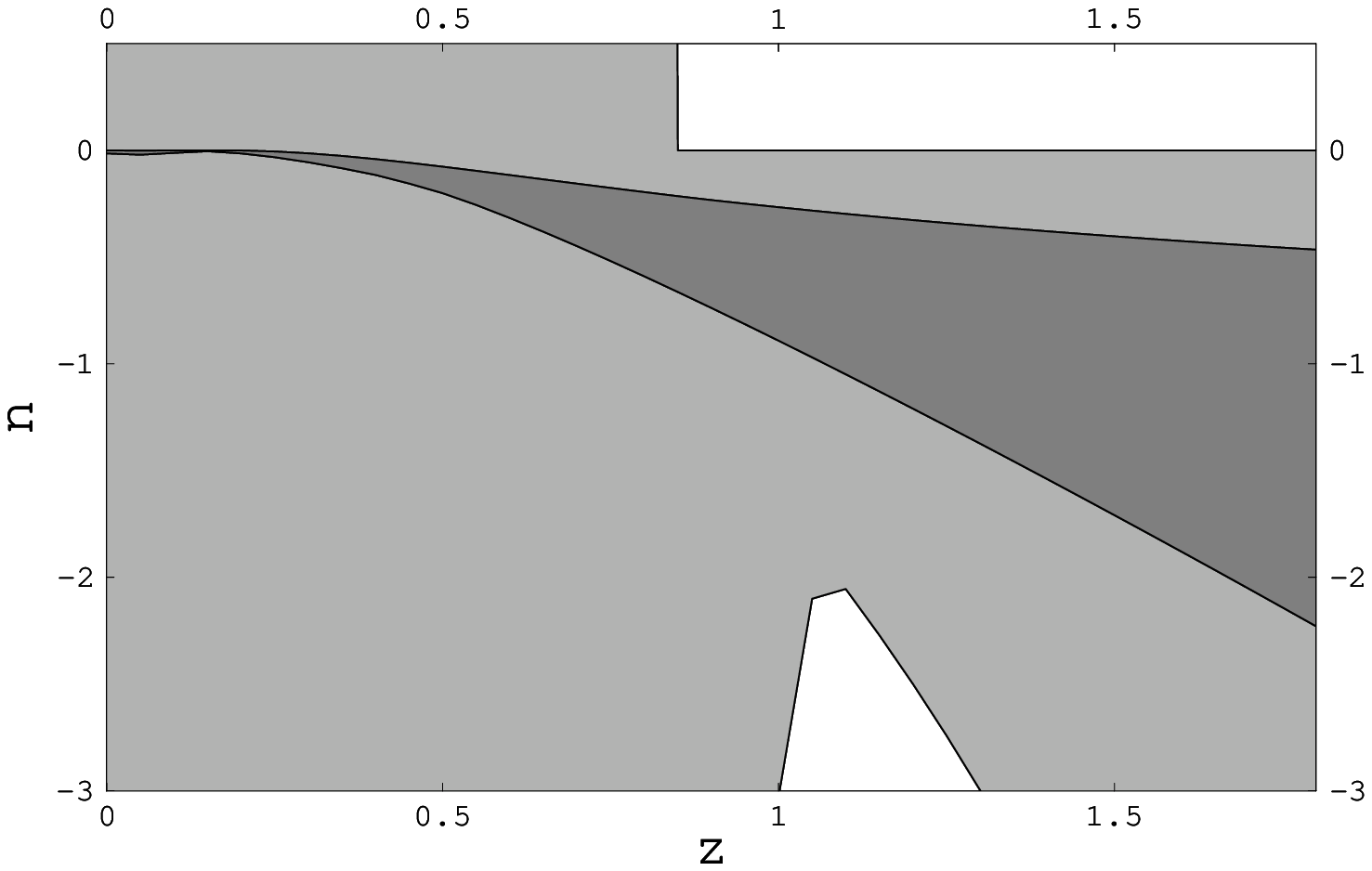}% Here is how to import EPS art
\caption{\label{fig:n-z} The characteristic of the power-law
potential reconstructed from data.}
\end{figure}

%%%%%%%%%%%%%%%%%%%%%%%%%%%% 2009.04 revised %%%%%%%%%%%%%%%%%%%%%%%%%%%%
In addition to the derivative, $dn(z)/dz$, the characteristic
$n(z)$ itself is also reconstructed from data, as illustrated in
Fig.\ \ref{fig:n-z} (dark gray for $68\%$ confidence and light
gray for $95\%$). According to the $95\%$ confidence constraint, a
negative power-law index between $-2$ and $0$ is favored, whereas
the model with a positive power is inconsistent with the current
data.
%%%%%%%%%%%%%%%%%%%%%%%%%%%%%%%%%%%%%%%%%%%%%%%%%%%%%%%%%%%%%%%%%%%%%%%%%
%In addition to the derivative of the characteristic, $dn(z)/dz$,
%the constraint on $n(z)$ itself is illustrated in Fig.\
%\ref{fig:n-z} (dark gray for $68\%$ confidence and light gray for
%$95\%$). According to the $95\%$ confidence constraint, a negative
%power-law index between $-2$ and $0$ is favored, whereas the model
%with a positive power is inconsistent with the current data.
%%%%%%%%%%%%%%%%%%%%%%%%%%%%%%%%%%%%%%%%%%%%%%%%%%%%%%%%%%%%%%%%%%%%%%%%%

%%%%%%%%%%%%%%%%%%%%%%%%%%%% 2009.05 added %%%%%%%%%%%%%%%%%%%%%%%%%%%%
\subsection{Distinguishing between models} \label{sec:III-D}
%%%%%%%%%%%%%%%%%%%%%%%%%%%%%%%%%%%%%%%%%%%%%%%%%%%%%%%%%%%%%%%%%%%%%%%

Here we demonstrate how well the two models considered above can be
distinguished from other models via our approach with the future supernova
observations. We invoke the SNAP-quality \cite{Aldering:2004ak} simulated
data with $2023$ SNe \cite{Shafieloo:2005nd} distributed in the redshift
range between $0$ and $1.7$, as well as the current-quality simulated CMB
\cite{WMAP:2008} and BAO \cite{Eisenstein:2005su} data. We take the other
models one by one as the fiducial model to generate $1000$ realizations of
the simulated data, with which we obtain the observational constraints on
$dQ_\textrm{exp}(z)/dz$ and $dQ_\textrm{power}(z)/dz$. If the observational
constraint is inconsistent with the theoretical prediction, $dQ(z)/dz=0$,
the fiducial model can be distinguished from the model with which the
characteristic $Q(z)$ is associated.

\begin{figure}
\begin{tabular}{cccc}
\scriptsize M1: $w_\textsc{de} = w_{\Lambda} = -1$ & % M1
\scriptsize M2: $w_\textsc{de} = -0.8$ & % M2
\scriptsize M3: $\displaystyle w_\textsc{de} = -1 + \frac{0.5z}{1+z}$ & % M3
\scriptsize M4: $\displaystyle w_\textsc{de} = -1 + \frac{1.5z}{1+z}$ % M4
\\ %
\hline %
\includegraphics[width=\figwidth]{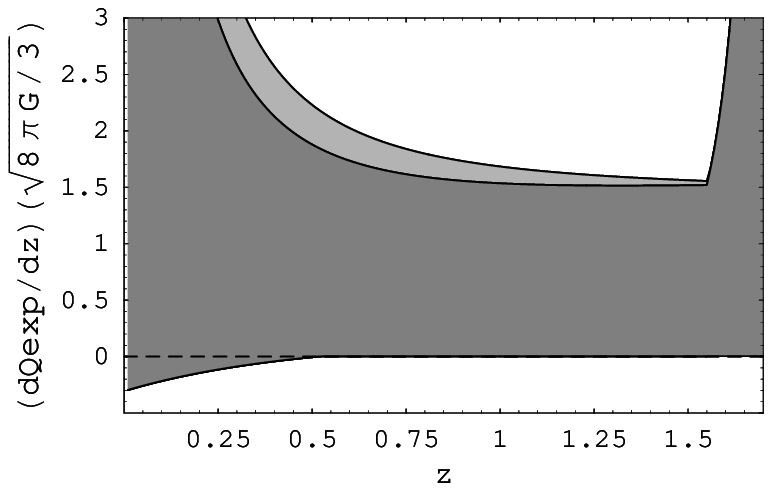} &
\includegraphics[width=\figwidth]{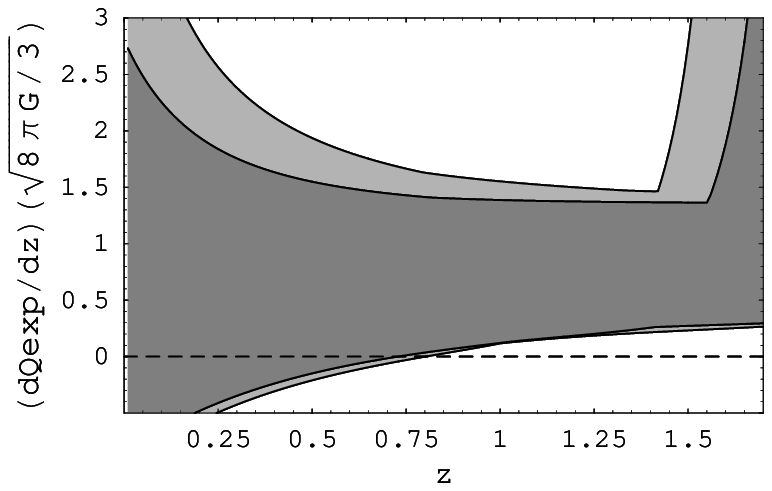} &
\includegraphics[width=\figwidth]{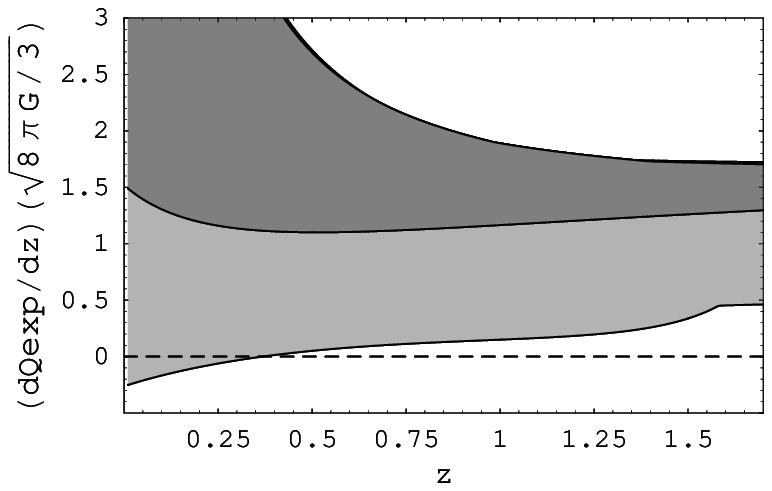} &
\includegraphics[width=\figwidth]{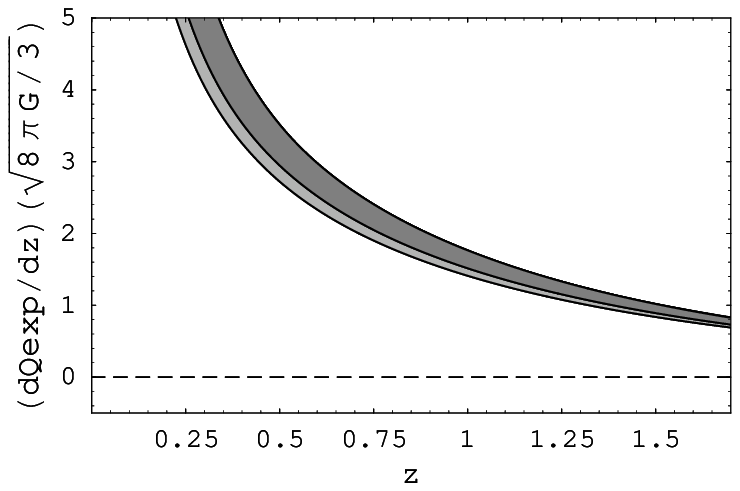} \\
\includegraphics[width=\figwidth]{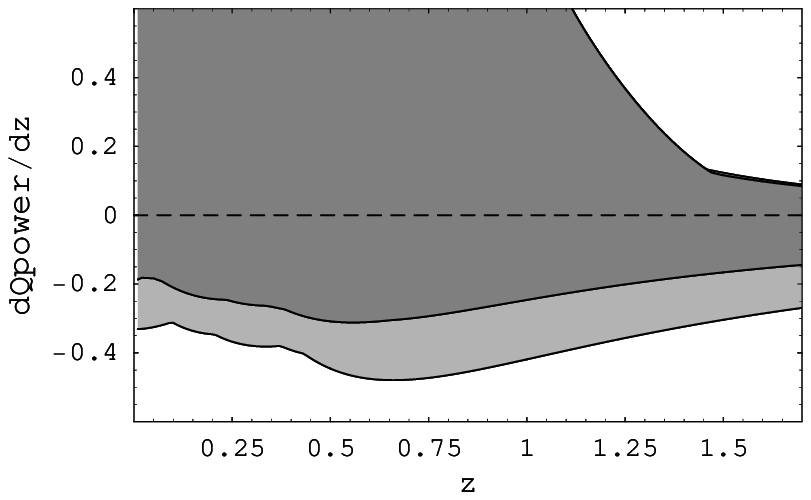} &
\includegraphics[width=\figwidth]{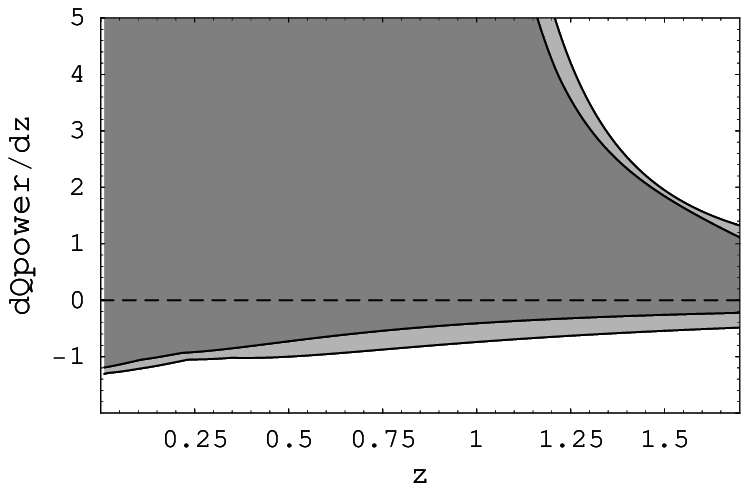} &
\includegraphics[width=\figwidth]{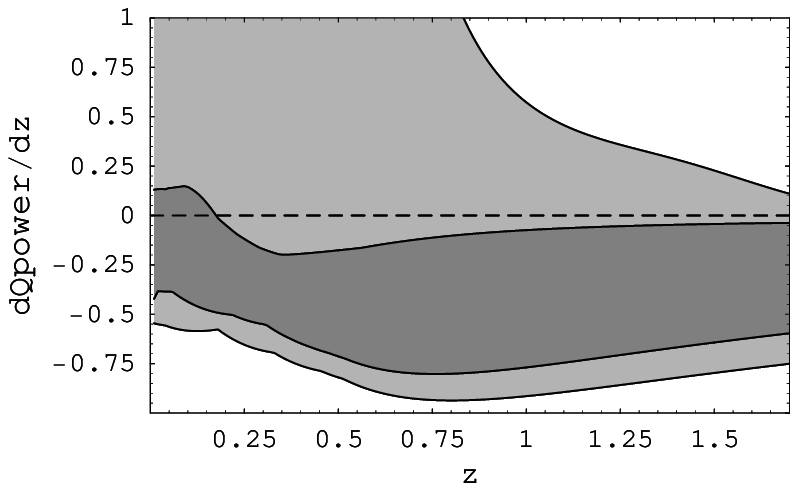} &
\includegraphics[width=\figwidth]{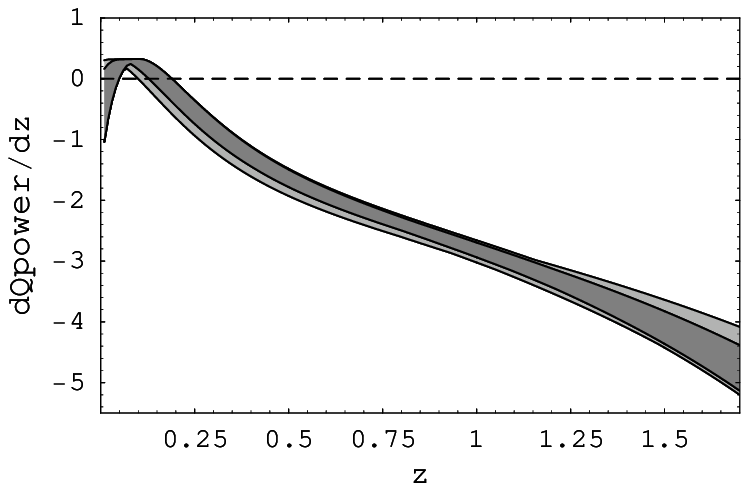} \vspace{0.5em}
\\ %
%\hline %
%\hline %
\scriptsize M5: $\displaystyle w_\textsc{de} = -0.8 - \frac{0.2z}{1+z}$ & % M9
\scriptsize M6: $\displaystyle w_\textsc{de} = -1.05 + \frac{0.2z}{1+z}$ & % M11
\scriptsize M7: $\displaystyle w_\textsc{de} = -0.6 - \frac{0.5z}{1+z}$ & % M13
\scriptsize M8: $\displaystyle w_\textsc{de} = -1.05 + \frac{1.0z}{1+z}$  % M10
\\ %
\hline %
\includegraphics[width=\figwidth]{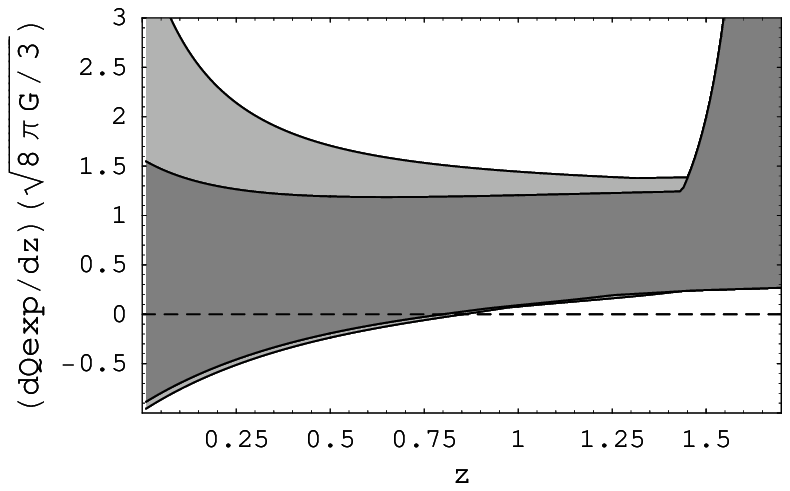} &
\includegraphics[width=\figwidth]{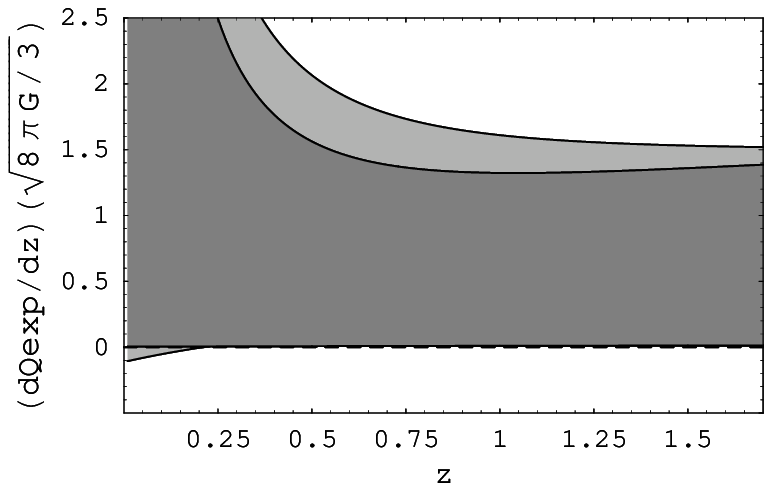} &
\includegraphics[width=\figwidth]{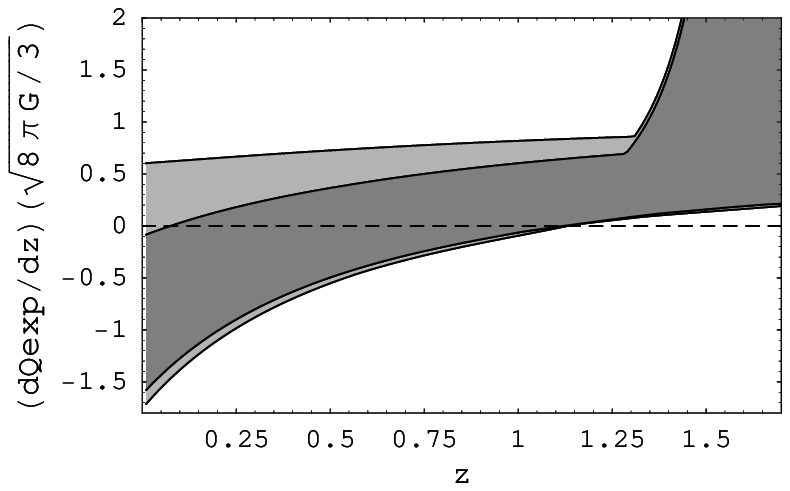} &
\includegraphics[width=\figwidth]{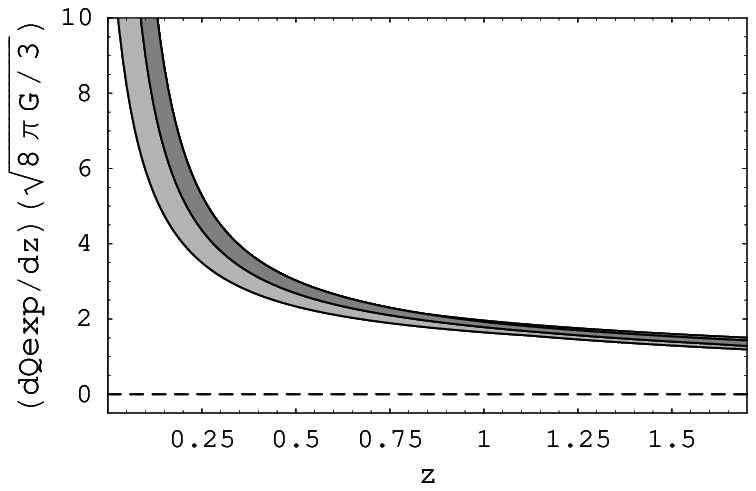} \\
\includegraphics[width=\figwidth]{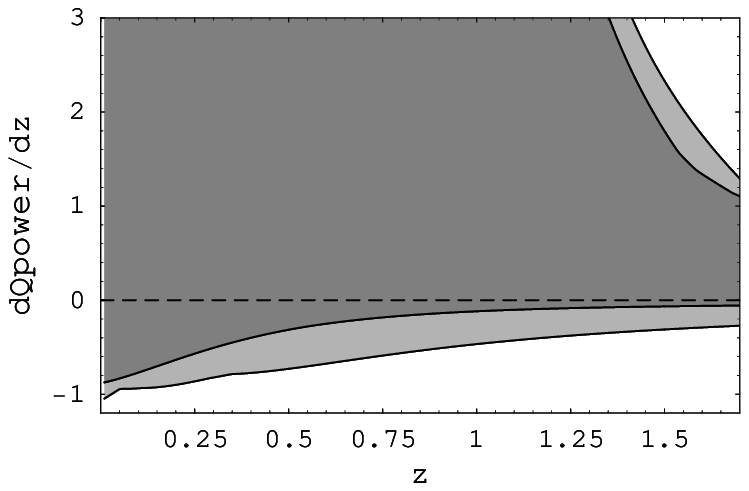} &
\includegraphics[width=\figwidth]{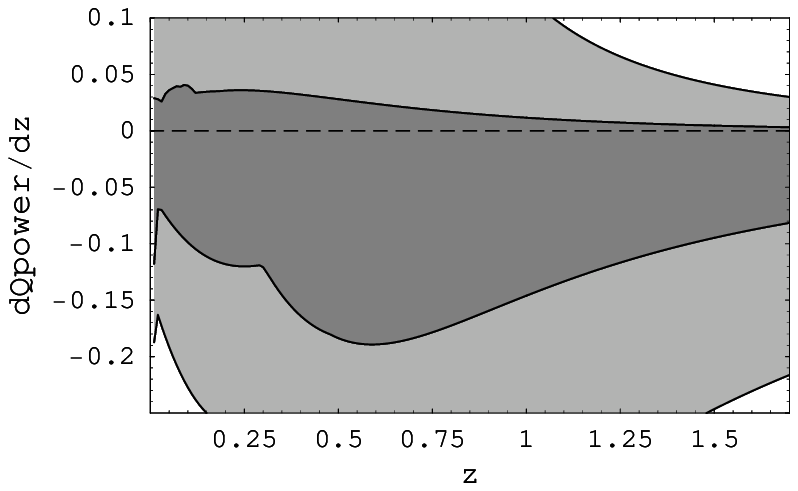} &
\includegraphics[width=\figwidth]{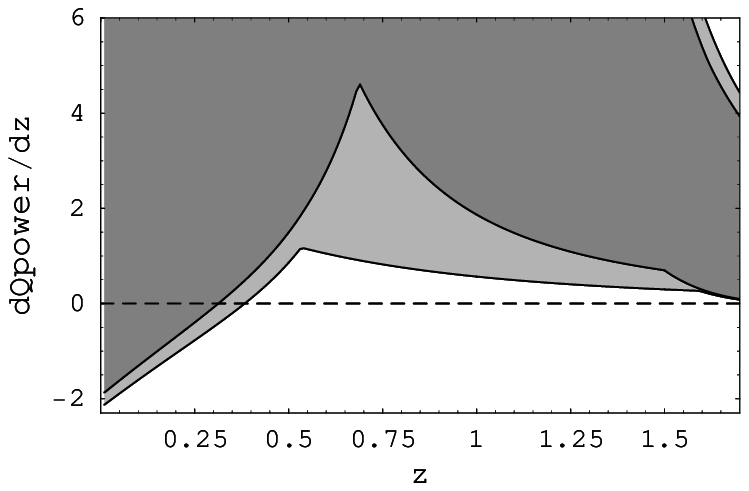} &
\includegraphics[width=\figwidth]{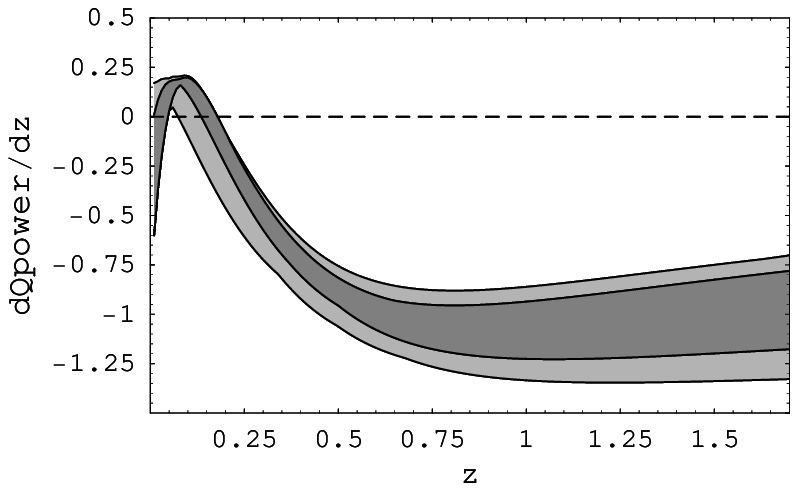}
\end{tabular}
\caption{\label{fig:fiducial} Observational constraints on
$dQ_\textrm{exp}(z)/dz$ and $dQ_\textrm{power}(z)/dz$ from the simulated
data generated w.r.t.\ eight fiducial modes.}
\end{figure}

We consider eight fiducial models: \\%
M1: \ $w_\textsc{de} = w_{\Lambda} = -1$, \\ % M1
M2: \ $w_\textsc{de} = -0.8$, \\% M2
M3: \ $w_\textsc{de} = -1 + 0.5z/(1+z)$, \\ % M3
M4: \ $w_\textsc{de} = -1 + 1.5z/(1+z)$, \\ % M4
M5: \ $w_\textsc{de} = -0.8 - 0.2z/(1+z)$, \\ % M9
M6: \ $w_\textsc{de} = -1.05 + 0.2z/(1+z)$, \\ % M11
M7: \ $w_\textsc{de} = -0.6 - 0.5z/(1+z)$, \\ % M13
M8: \ $w_\textsc{de} = -1.05 + 1.0z/(1+z)$. \\ % M10
(The first four models are considered in \cite{Zunckel:2008ti}.) The
observational constraints obtained w.r.t.\ these fiducial models on
$dQ_\textrm{exp}(z)/dz$ and $dQ_\textrm{power}(z)/dz$ are presented in
Fig.\ \ref{fig:fiducial} (dark gray for $68\%$ confidence and light gray
for $95\%$). As shown by this figure, with the future observations
considered above and via our approach, the exponential-potential
quintessence model is distinguishable from all the eight dark energy
models, while the power-law potential can be distinguished from the models
with faster evolving $w$: M3 ($68\%$ confidence), M4, M7 and M8 ($95\%$
confidence), but not from those with more slowly evolving $w$: M1, M2, M5
and M6.

%%%%%%%%%%%%%%%%%%%%%%%%%%%%%%%%%%%%%%%%%%%%%%%%%%%%%%%%%%%%%%%%%%%%%%%
\section{Summary} \label{sec:summary}
%%%%%%%%%%%%%%%%%%%%%%%%%%%%%%%%%%%%%%%%%%%%%%%%%%%%%%%%%%%%%%%%%%%%%%%

%%%%%%%%%%%%%%%%%%%%%%%%%%%% 2009.04 revised %%%%%%%%%%%%%%%%%%%%%%%%%%%%
In this paper we propose a new approach to the testing and the
constraining of dark energy models based on observational results.
To demonstrate our approach, we concentrate on quintessence and
proceed in three directions: (A) We reconstruct the quintessence
potential. (B) We assess the consistency between quintessence
models and data. (C) We obtain the constraint on the distinct
parameter of one category of quintessence models.
%%%%%%%%%%%%%%%%%%%%%%%%%%%%%%%%%%%%%%%%%%%%%%%%%%%%%%%%%%%%%%%%%%%%%%%%%

For Direction B, to assess the consistency between quintessence
models and data, we introduce a characteristic $Q(z)$ for each
category of theoretical potentials. This characteristic $Q(z)$ is
in general time-varying, but within the scope of those potentials
it is constant and equivalent to a distinct parameter therein. By
comparing the reconstructed $dQ(z)/dz$ with zero we can assess the
consistency between data and that category of potentials. This
approach provides a simple % and direct
``litmus test'' for each category of quintessence models. For
Direction C, since the characteristic $Q(z)$ plays the role of the
distinct parameter within the scope of a category of models, via
reconstructing $Q(z)$ from data we obtain the constraint on that
distinct parameter.

%%%%%%%%%%%%%%%%%%%%%%%%%%%% 2009.04 revised %%%%%%%%%%%%%%%%%%%%%%%%%%%%
Our approach to the consistency test and the constraining of dark
energy models is simple and efficient to perform. This is because
in this approach (i) for all models the characteristics $Q(z)$ and
their derivatives $dQ(z)/dz$ are reconstructed from data via the
observational constraints of a single parameter space that by
choice can be easily accessed, and (ii) with our design of the
characteristic $Q(z)$ we can test the models and constrain their
distinct parameter without the knowledge of the other parameters
of the models. The simplicity and the efficiency of our approach
is particularly manifest when it is applied to quintessence. In
this case, in addition to the above-mentioned two general
features, a specific technical reason for this benefit is that in
our approach one does not need to numerically solve the
quintessence field equation and the Einstein equations coupled
together.
%%%%%%%%%%%%%%%%%%%%%%%%%%%%%%%%%%%%%%%%%%%%%%%%%%%%%%%%%%%%%%%%%%%%%%%%%
%Our approach to the consistency test and the constraining of dark
%energy models is particularly simple and efficient to perform.
%This is because in this approach (i) for all models the
%characteristics $Q(z)$ and their derivatives $dQ(z)/dz$ are
%reconstructed from data via the observational constraints of a
%single parameter space that by choice can be easily accessed, and
%(ii) with our design of the characteristic $Q(z)$ we can test the
%models and constrain their distinct parameter without the
%knowledge of the other parameters of the models. In particular,
%when applied to quintessence, our approach is significantly
%simpler and faster than the model-based approach. In addition to
%the above-mentioned two general features, a particular technical
%reason for the high efficiency of our approach is that we do not
%need to numerically solve the quintessence field equation and the
%Einstein equations, which has to be done repeatedly in the
%model-based approach and can be time-consuming.
%%%%%%%%%%%%%%%%%%%%%%%%%%%%%%%%%%%%%%%%%%%%%%%%%%%%%%%%%%%%%%%%%%%%%%%%%

To demonstrate the effectiveness of our approach, we invoked a
widely used % model-independent
parametrization of the dark energy EoS, and investigated the
exponential potential and the power-law potential of quintessence
models. For each of them a characteristic was introduced. Via
these characteristics, the two theoretical potentials were
compared with the current data. We found that at the $68\%$
confidence level both the exponential and the power-law potentials
were ruled out. When relaxed to the $95\%$ confidence constraint,
the power-law potential with a negative power-law index between
$-2$ and $0$ can fit the current data. In contrast, the
exponential potential remains disfavored.

%%%%%%%%%%%%%%%%%%%%%%%%%%%% 2009.04 revised %%%%%%%%%%%%%%%%%%%%%%%%%%%%
To perform our approach, one may choose other parametrizations.
%This manner of testing and constraining models is as
%model-independent as the chosen parametrization.
Generally speaking, an approach invoking parametrization may be
accompanied with a bias against certain models. This is an
essential issue that requires further study.
% The issue about the model-independence and
The potential bias of performing our approach via the
parametrization of the dark energy EoS in Eq.\
(\ref{eq:w0wa-parametrization}) is currently under our
investigation \cite{Chen:2009}.
%%%%%%%%%%%%%%%%%%%%%%%%%%%%%%%%%%%%%%%%%%%%%%%%%%%%%%%%%%%%%%%%%%%%%%%%%

%%%%%%%%%%%%%%%%%%%%%%%%%%%% 2009.04 revised %%%%%%%%%%%%%%%%%%%%%%%%%%%%
To sum up, our approach provides a useful tool for testing and
constraining dark energy models based on observational results.
This approach can be applied to a variety of quintessence models
and other dark energy models, and probably to other explanations
of the cosmic acceleration. The general principle of our approach
may also be applied to different cosmological models and even the
models in other fields beyond the scope of cosmology.
%%%%%%%%%%%%%%%%%%%%%%%%%%%%%%%%%%%%%%%%%%%%%%%%%%%%%%%%%%%%%%%%%%%%%%%%%

%%%%%%%%%%%%%%%%%%%%%%%%%%%%%%%%%%%%%%%%%%%%%%%%%%%%%%%%%%%%%%%%%%%%%%%%%
\begin{acknowledgments}
Gu is supported by the National Center for Theoretical Sciences
(funded by the National Science Council), Taiwan, R.O.C.,
C.-W. Chen by the Taiwan National Science Council under Project
No. NSC 95-2119-M-002-034 and NSC 96-2112-M-002-023-MY3,
and P. Chen by Taiwan National Science Council under Project No.
NSC 97-2112-M-002-026-MY3 and by US Department of Energy under
Contract No. DE-AC03-76SF00515.
\end{acknowledgments}
%%%%%%%%%%%%%%%%%%%%%%%%%%%%%%%%%%%%%%%%%%%%%%%%%%%%%%%%%%%%%%%%%%%%%%%%%

\newpage

%%%%%%%%%%%%%%%%%%%%%%%%%%%%%%%%%%%%%%%%%%%%%%%%%%%%%%%%%%%%%%%%%%%%%%%%%%

%%%%%%%%%%%%%%%%%%%%%%%%%%%%

%%%%%%%%%%%%%%%%%%%%%%%%%%%%

%%%%%%%%%%%%%%%%%%%%%%%%%%%%

%%%%%%%%%%%%%%%%%%%%%%%%%%%%

%%%%%%%%%%%%%%%%%%%%%%%%%%%%%%%%%%%%%%%%%%%%%%%%%%%%%%%%%%%%%%%%%%%%%%%%%%

\end{document}